\documentstyle[twocolumn,aps,psfig]{revtex}

\begin{document}
\draft

\twocolumn[\hsize\textwidth\columnwidth\hsize\csname
@twocolumnfalse\endcsname

\title{Pseudo-gap behavior in dynamical properties\\
of high-$T_c$ cuprates}

\author{T\^oru Sakai and Yoshinori Takahashi}
\address{
Faculty of Science, Himeji Institute of Technology, Kamigori,
Ako-gun, Hyogo 678-1297, Japan
}

\date{June 98}
\maketitle

\begin{abstract}
Dynamical properties of 2D antiferromagnets with hole doping
are investigated to see the effects of short range local magnetic order
on the temperature dependence of the dynamical magnetic susceptibility.
We show the pseudo-gap like behavior of the temperature dependence of
the NMR relaxation rate. We also discuss implications of the results in
relations to the observed spin gap like behavior of low-doped copper
oxide high-$T_c$ superconductors.
\end{abstract}

\vskip2pc]
\narrowtext

The spin-gap-like behavior observed in some low-doped high-$T_{\rm c}$
cuprates by dynamical measurements\cite{nmr1}
have recently attracted much 
interest.
The term ``gap'' originally arose because of the appearance
of a broad peak above $T_{\rm c}$ in the temperature dependence of the NMR
relaxation time measurements. It is, however, not clear whether the
gaplike
behavior is really associated with the opening of the energy gap in the
excitation spectrum of the system, because the presence of 
superconductivity does not allow 
the measurement of the temperature-dependent
properties at low temperature.

There have been several theoretical explanations for the origin of
the gaplike behavior, for instance, 
a spinon singlet formation mechanism\cite{spinon} and a CuO$_2$
bilayer coupling mechanism.\cite{bilayer1,bilayer2} Its origin
is, however, still controversial and is debatable. The purpose of
the present paper is to propose a new theoretical explanation for the
origin of the observed pseudogap behavior of cuprate superconductors.

Our idea is based on the observation that the pseudogap behavior
is characteristic of low-doped cuprates, including, in particular, the
pure
undoped 2D Heisenberg antiferromagnets as an extreme case.
We also assume that it is a property already present in the single-
layer system, noting the fact that the gaplike behaviors are observed
even in single-layer high-$T_{\rm c}$ cuprates.\cite{nmr2}
It is then quite natural
to assume that the observed pseudogap behavior will arise as a
result
of the short-range magnetic order (SRO) of the
low-dimensional Heisenberg antiferromagnets with lowering of the
temperature. 
The gaplike behavior has also been discussed in associating with the
antiferromagnetic correlations in terms of the
spin fluctuation theory.\cite{miyake} 
Our purpose is to test this concept 
by explicit numerical diagonalization of the $t$-$J$ model.

In low-dimensional magnets, even if there appears no long-range magnetic
order, SRO begins to grow when the
temperature of the system decreases below the characteristic
temperature, 
which is of the same magnitude as the magnetic coupling constant.
The excitation spectrum of the system is then affected by the
presence of the antiferromagnetic SRO, i.e., spin excitations are
suppressed,
resulting in the deviation of the Curie-Weiss-like temperature
dependence of
$1/T_1T$ as well as giving rise to a broad peak around a cross-over
temperature. In this way, without assuming the real
energy gap, we are still able to explain the observed gaplike
temperature
dependence.  By hole doping, SRO will be
rapidly suppressed, which will also be in accordance with the observed
hole-concentration dependence of the spin-gap temperature $T_{\rm s}$.

\begin{figure}
\begin{center}
\mbox{\psfig{figure=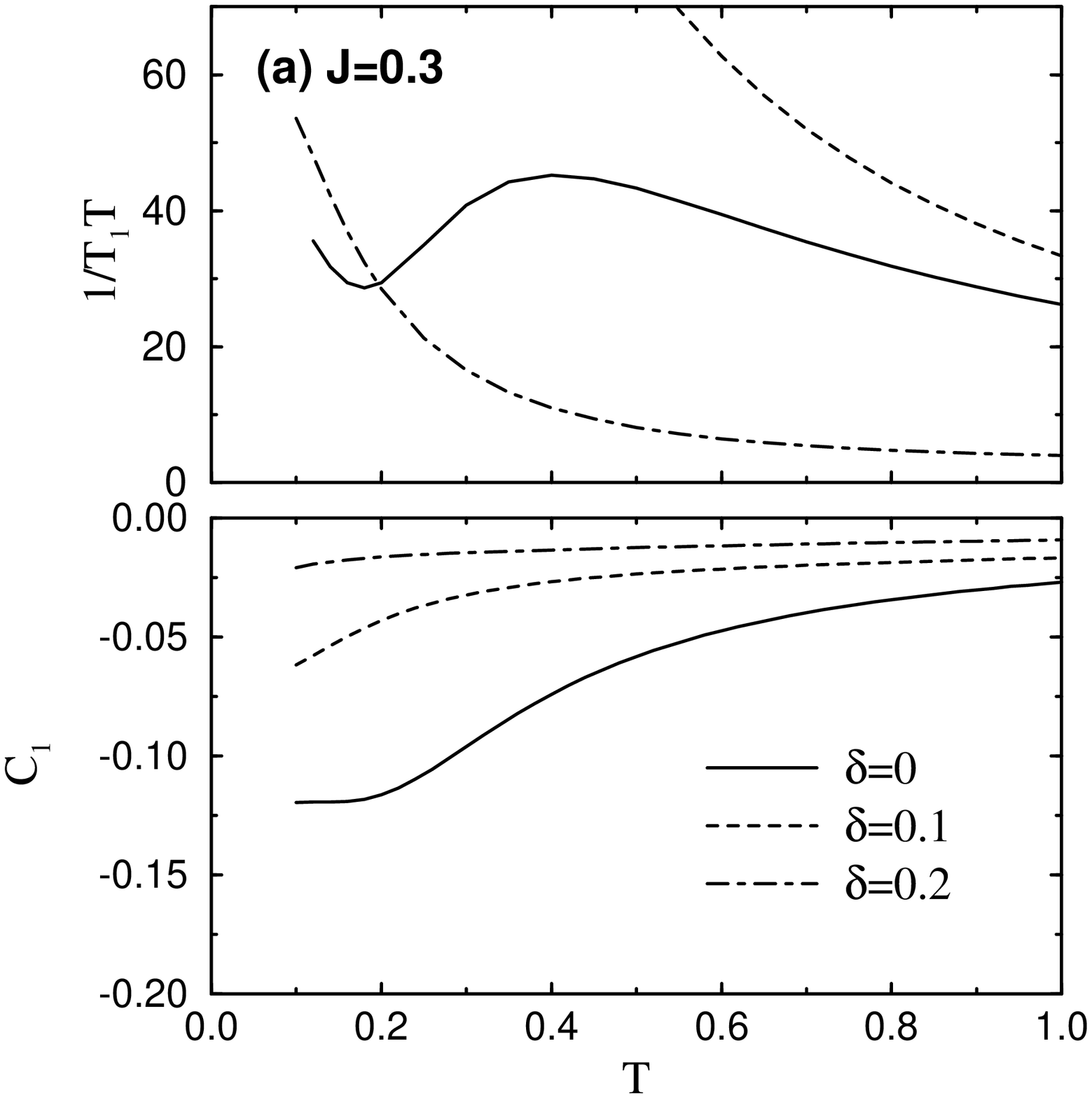,width=8cm,height=8cm,angle=-90}}
\mbox{\psfig{figure=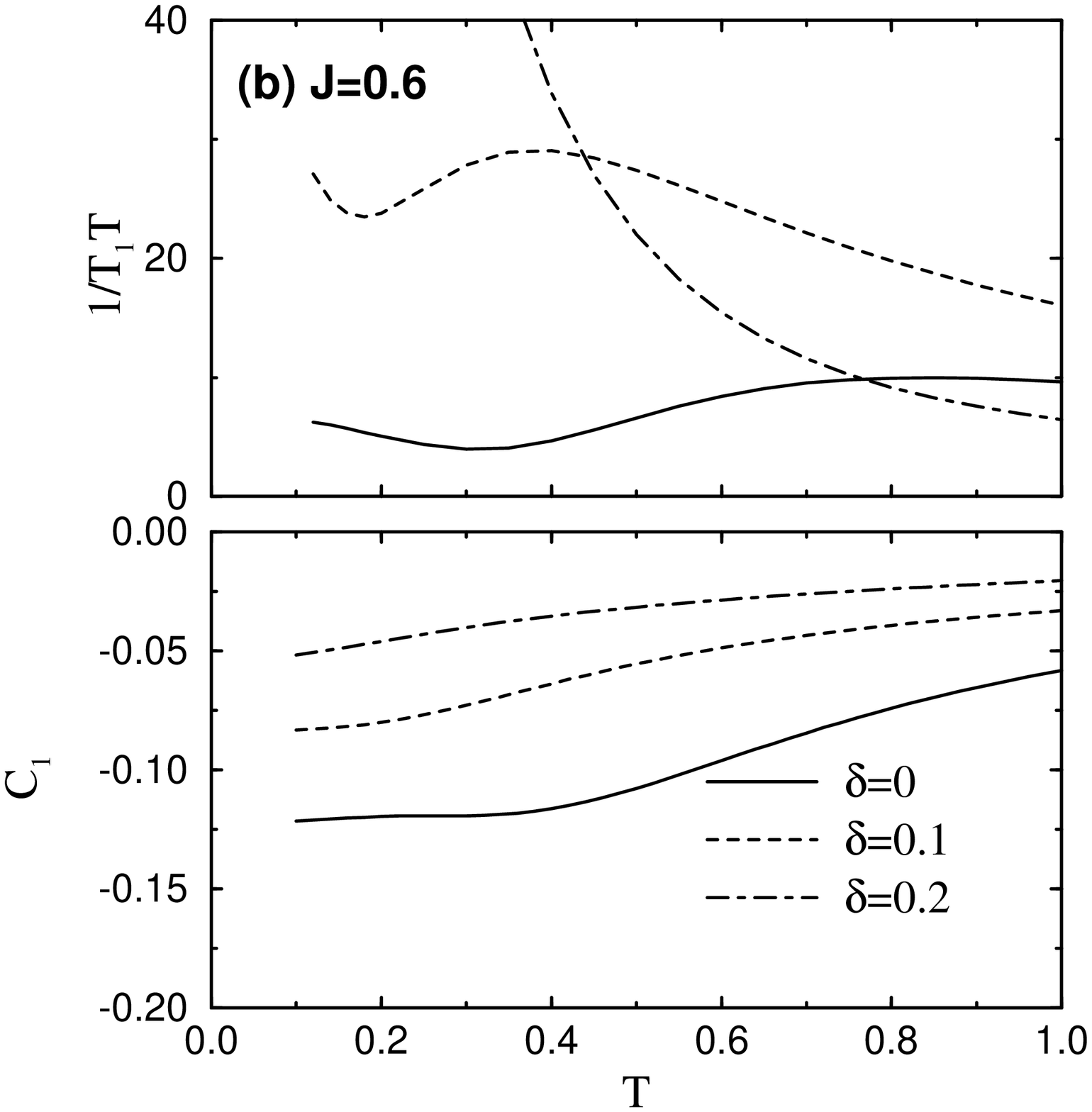,width=8cm,height=8cm,angle=-90}}
\end{center}
\caption{
Temperature dependence of $1/T_1T$ (the right hand side of
(2) in the text) and NN spin correlation $C_1$ for (a) $J=0.3$ and
(b) $J=0.6$.
}
\label{fig1}
\end{figure}

The Hamiltonian of the present study is the finite-size single-layer 
$t$-$J$ model given by 
\begin{equation}
\label{ham}
H = - t \sum_{<\bf i,\bf j>, \sigma}
            ( {c}_{\bf j,\sigma}^\dagger {c}_{\bf i,\sigma}
           + {c}_{\bf i,\sigma}^\dagger {c}_{\bf j,\sigma} )
  + J \sum_{<\bf i,\bf j>}
           ( {\bf S}_{\bf i} \cdot {\bf S}_{\bf j}
           - \textstyle{1 \over 4} n_{\bf i} n_{\bf j} ) ,
\end{equation}
where $t$ is the nearest-neighbor (NN)
electron hopping integral and $J$ is
the antiferromagnetic Heisenberg exchange constant between spins on
adjacent lattice sites. Throughout the paper, all the energies are
measured in units of $t$. With the use of the calculated eigenvalues and
eigenvectors, we evaluated the temperature dependence of
the NMR relaxation rate $1/T_1$
by the following formula:\cite{calculation}
\begin{equation}
\label{t1t}
{1\over {T_1T}} \propto \lim_{\omega \to 0}{1\over {\omega}}\sum_q {\rm
Im}
\chi (q,\omega) ,
\end{equation}
where ${\rm Im}\chi (q,\omega)$ is the imaginary part of the
dynamical spin susceptibility of conduction electrons.
The effect of the $q$-dependence of the hyperfine form factor
is neglected, for simplicity.
In actual numerical estimations of ${\rm Im}\chi (q,\omega)$
the $\delta$-function is approximated by the Lorentzian distribution
with a small width $\epsilon =0.01t$.
In order to reveal the relation between the temperature dependence of the
relaxation rate and SRO, the NN spin correlation
function is also calculated by
\begin{equation}
\label{cor}
C_1={1\over {N}}\sum _{i}{1\over 4}
\sum_{\rho=\pm \hat x,\pm \hat y}\langle S^z_i S^z_{i+\rho} \rangle .
\end{equation}
Since we need all the eigenvalues and eigenvectors, the cluster size of
the model was limited by the available disk space of the computer.
We show, in Fig.\ref{fig1}(a), the temperature dependence of $1/T_1T$ and
NN spin correlation function for the $\sqrt {10} \times \sqrt {10}$
cluster
with 0, 1 and 2 holes, corresponding to the hole concentrations,
$\delta$=0, 0.1 and 0.2, respectively.
We employ $J=0.3$, estimated for cuprate superconductors.
We clearly see a broad peak in the temperature dependence of 
$1/T_1T$
of the undoped system around the temperature $T\sim J$, which agrees
with the temperature below which SRO begins to prevail.

Unfortunately no significant suppression of $1/T_1T$ is observed in
systems with hole dopings, as shown in the same figure. This is,
however, consistent with the absence of SRO in these systems.
For small clusters, the relative importance of the effect of electron
hopping tends to be overestimated for the fixed value of $J$.
Therefore we also performed the same calculation assuming
a larger parameter value of $J$ (=0.6). From the $T$-dependence of
$1/T_1T$ in Fig.\ref{fig1}(b) with $\delta=0.1$, we clearly see that
the gaplike behavior is also present even in the doped system, which 
correlates well with the appearance of SRO. From the $\delta$-dependence
of the figure, we also see that the temperature region having SRO rapidly
decreases with increasing hole concentration.
Since the calculated lowest excitation gap of
a $\sqrt {10} \times \sqrt {10}$ cluster with one hole is 0.146,
the broad peak around $T\sim 0.4$ is not related to 
any finite size effects.
We have confirmed that the same correlation between that $1/T_1T$ and SRO
also exists on the $\sqrt {8} \times \sqrt {8}$ cluster with one
hole ($\delta$=0.125), which suggests that the feature is a bulk property
independent of the system size.
The same calculation with hole dopings was reported for  
the $4\times 4$ cluster\cite{jaklic} ($J/t=0.3$). However, the temperature
range did not cover the region where the peak of $1/T_1T$ was expected
to be observed.
We have confirmed that the broad peak structure in the $T$-dependence
of $1/T_1 T$ is almost independent of the Lorentzian width
$\epsilon$. The absolute values of $1/T_1 T$, on the other hand,
are very sensitive to our choice of $\epsilon$ 
in such small cluster calculations. Therefore the relative
magnitude of the curves in Fig.\ref{fig1}, evaluated by assuming the same
$\epsilon$ value for several $\delta$ values, should not be taken
seriously. For the same reason, $1/T_1 T$ has a large but
finite value at low temperature for $\delta=0.2$.

In the Heisenberg model with no hole doping, the presence of
the broad peak in the temperature dependence of $1/T_1T$ around $T\sim
J$
has already been derived by Chakravarty and Orbach\cite{chakravarty}
based on the high-temperature series-expansion method.
They also predicted that 
$1/T_1T$ shows a minimum in its $T$-dependence 
around $T\sim J/2$, which was later verified by the quantum
Mont-Carlo calculation\cite{sandvik} for the case of local hyperfine
coupling. The upturn behavior of $1/T_1T$ for $\delta =0$ in our
Figs.\ref{fig1}(a) and \ref{fig1}(b) may arise by the above mechanism.
Owing to the finite-size effects
we must, however, be careful when drawing any definite
conclusions concerning 
whether these minima will survive even in the presence of
hole doping in the thermodynamic limit at low temperature.
In bulk two-dimensional Heisenberg systems without hole doping, the
broad peak at around $T\sim J$ may be difficult to observe as clearly
as in Fig. \ref{fig1}, 
because the long-range antiferromagnetic order grows
toward $T=0$ 
at low temperature. 
A modified high-temperature series
expansion study\cite{singh} actually yielded not 
such a clear peak, but a small shoulder in the temperature
dependence of $1/T_1T$. Experimentally, though no peak structure in
the $T$ dependence of $1/T_1T$ for La$_2$CuO$_4$ has been observed by
Imai {\it et al.},\cite{imai} its absence is still controversial
because the temperature range is limited to below $T\sim J$.
On the other hand, in the case of systems with hole doping,
we expect 
the peak behavior to be more easily observed than in the pure Heisenberg
system, due to the absence of the antiferromagnetic long-range order
in the ground state at $T=0$ K.

In conclusion, we have succeeded in deriving
the pseudogap behavior of high-$T_{\rm c}$ cuprates.
We showed that a broad peak appears in the temperature
dependence of $1/T_1T$ because of the suppression of the dynamical
susceptibility due to the development of SRO
inherent to undoped low-dimensional Heisenberg
antiferromagnets. It will be interesting to examine the effect of SRO on
other dynamical quantities, such as neutron scattering intensities and
the photoemission spectrum. 
We could also show that hole doping rapidly
destroyed the SRO of the system, resulting in the observed
hole-concentration dependence of the spin-gap temperature.

If the present scenario is true, the hole concentration dependence of
the superconducting critical temperature $T_{\rm c}$ can be understood
as follows. First the behavior seems to suggest 
that the antiferromagnetic SRO 
is necessary for the superconducting mechanism of 
high $T_{\rm c}$ cuprates to work.
In the underdoped region, because the SRO is always present since  
$T_{\rm s} > T_{\rm c}$, $T_{\rm c}$ is
mainly determined by some pairing mechanism, giving rise to the
critical temperature proportional to the hole concentration.
On the other hand, in the overdoped
region, though the mechanism might indicate a higher $T_{\rm c}$,
it is upper-limited by $T_{\rm s}$ because of the disappearance of
SRO. We are not concerned with the nature of the pairing mechanism
of the superconductivity in the present paper. The presence of SRO,
in this way, may play a significant role as a necessary condition
for the occurrence of the high-$T_{\rm c}$ superconductivity.

We wish to thank Prof. Y. Hasegawa for fruitful
discussions.
We also thank
the Supercomputer Center, Institute for
Solid State Physics, University of Tokyo for the facilities
and the use of the Fujitsu VPP500.

\end{document}